# Image processing Application Development on Software Configurable Processor Array

Ganesh Prabhu[#1], Steevan Rodrigues[*2], Niranjan Chiplunkar[#3]  Niranjan U.C[*4,]

[#]CSE, VTU
N.M.A.M.I.T Nitte Karkala, Udupi,India
[1]ganeshsprabhu@gmail.com
[3]niranjanchiplunkar@rediffmail.com

[*]Manipal Dot Net PVT Ltd.
Manipal, India
[2]steevan.rodrigues@manipal.net
[4]niranjan@manipal.net

*Abstract*—The software configurable processor finds best use in the embedded systems. These processors have on-chip logic like FPGA (Field Programmable Gate Array) and thus can be configured to implement custom hardware functionality. The digital computing tasks that need to accelerate in hardware can be compiled down to a configuration file or bit stream that contains the information on how the logic components are configured and wired together.
Video and image processing applications perform repeated pixel transformation and thus consume more power and the processing time. Software configurable processor best accelerates such compute intensive applications. This paper mainly  focuses on the implementation of an image processing application called median filtering on a single processor and colour conversion algorithm on array of SCPs(Software Configurable Processors). Median filtering result on Digital Video Recorder (DVR) is discussed as a real-time application of SCP.

## I. Introduction

Software configurable processors are mainly used to implement compute intensive applications. The compute intensive applications do repetitive arithmetic operations on data samples and thus consume large amount of computation time. The compute intensive part of an application is identified by utility profiling [14] and can be accelerated on the SCP. To develop the application for the SCP, the programmer identifies critical sections to be accelerated, writes one or more extension instructions [10] as functions in C programming language, and accesses those functions from the application program. Performance gains of more than an order of magnitude over the un-accelerated processor can be achieved through vector processing [14].

Stretch provides a parallel processing platform called Software Configurable Processor Array (SCPA).  This array is made of multiple SCPs, wherein one of them acts as master processor interacting with the external world [3]. Stretch provides a very user-friendly set of application level programming interfaces for programming the multiple processors. The application can be run as number of tasks running simultaneously on these processors. Different colour conversion algorithms such as RGB-YIQ, RGB-YIQ RGB-YUV and RGB-CMY [12], [11],[14] are run on the multiple SCPs in parallel. This speeds up the application.

Median filter algorithm for removing impulse noise is implemented on ISEF (Instruction Set Extension Fabric) and the same algorithm is run on DVR. This algorithm is very widely used in digital image processing because under certain conditions it preserves the edges while removing noise. Implementation of the median filter algorithm is discussed in this paper.

DVR for the security began with the mechanical type of devices, they were operated by remote controllers or buttons on DVR box itself. Today both PC based and embedded DVRs are available in the market. The DVR used in the work reported here, is based on software configurable processor and it is widely used in video surveillance industry [5]. This multi-channel video recorder supports different video decimations and video standards. Latest compression technology is supported along with the older technologies [6].

Colour conversion algorithm is explained as an application of configurable processor array. The multi-channel video recorder in real time uses the multiprocessor environment for encoding and decoding channels. A set of channels are handled by different processors. The DVR find its use in banking sectors, shopping malls and other places where the security is the main concern.

## II. SCPA ARCHITECTURE

The S6000 is a family of SCP of Stretch Inc [9], [10] whose processor array connection is shown Fig.1

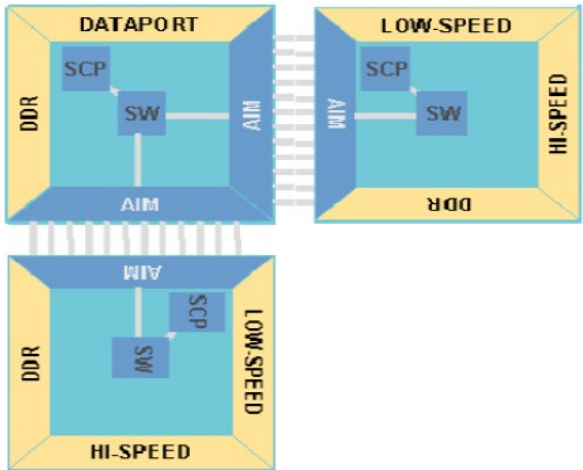

Fig. 1 Device connection using processor array

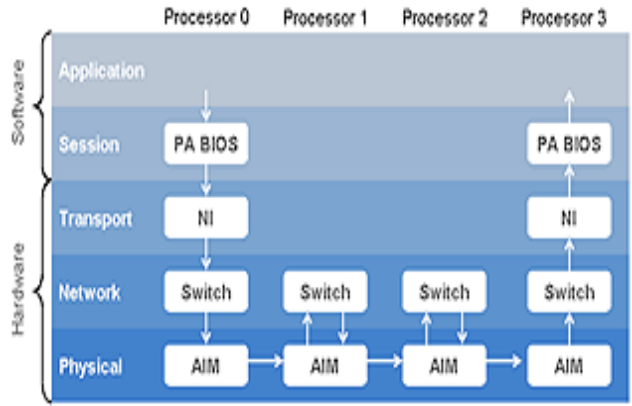

Fig. 2 The processor array network implementation

The SCP has a core made up of Tensilica LX, which is VLIW RISC architecture. The software-configurable part of the processor is ISEF, which is capable of doing multiple arithmetic and logical operations in parallel. The ISEF consists of multiple multipliers, arithmetic units, registers and multiplexers; and is embedded in the processor. The S6000 for example is made up of 64 multiplication units each capable of 8x16bit multiplication, 4096 arithmetic units and 64KB of embedded RAM [14].

In applications where extreme performance is required, multiple processors may be needed to achieve the required compute bandwidth. With conventional microprocessors and DSPs, this poses a challenge of partitioning the application as well as connecting the devices and arbitrating among them. The S6000 family of devices overcomes this challenge with its Processor Array. The goal of the Processor Array interface is to abstract away the inter-chip communication and allow multiple devices to collaborate and be designed in a single cohesive integrated development environment.

The S6000 family Processor Array technology consists of a set of BIOS calls (PA-BIOS), a physical interface called the Array Interface Module (AIM), and a Network Interface and Switch. Each processor in the S6000 family has four AIM ports that are designed to communicate at high speed and to connect together without glue logic. Each port can move data between devices at a rate of over 2.4GB/s (1.2GB/s in each direction). The processor array network implementation is shown in the Fig.2

## III. IMPLEMENTATION OF MEDIAN FILTER ALGORITHM ON SCP

Image processing algorithms when implemented on software configurable processor improves the performance of the image processing system. Median filter is a non-linear digital filter, which is able to preserve sharp signal changes, and is very effective in removing impulse noise (or salt and pepper noise). An impulse noise has either higher or lower Gray level value that is different from the neighbourhood pixels.

During the early days of development of digital image processing techniques, linear techniques were used extensively because of their mathematical simplicity and the existence of appropriate characteristics (e.g. principle of superposition) making them easy to design and implement. In the case where noise can be modulated as additive Gaussian noise, linear techniques offer satisfactory performance for noise removal. However, in many cases the noise is impulsive and in such a case linear techniques do not usually perform well. Another example where linear techniques fail is the case of non-linear image degradations. Such degradations occur during image formation and during image transmission through non-linear channels.

Linear filters do not have the ability to remove this type of noise without affecting the distinguishing characteristics of the signal. Median filters have remarkable advantages over linear filters for this particular type of noise [7]. Therefore, median filter is very widely used in digital signal and image/video processing applications [4].

A median filter works by setting, in turn, the value of each pixel in an image (except for the pixels on the border) to the median of the values of the pixels in a window surrounding the pixel. Median filters can be used to remove scattered noise from images and smooth them [1], while preserving the edges of objects in the image. There are different ways to deal with the pixels on the border of the image. In this implementation, the values of those pixels are padded with zeros.

A median filter is implemented by sliding a window of odd size over an image. A windowed median operation is implemented by sliding a window of odd size (e.g. 3x3 window) over an image. At each window position the sampled values of signal or image are sorted, and the median value of the samples replaces the sample in the centre of the window.

The main problem of the median filter [8] is its high computational cost (for sorting 3x3 pixels), even with the most efficient sorting algorithms. In the present work, sorting operation is performed in the ISEF with help of wide register. Nine pixels are packed into WR and sent to the ISEF, which sorts them and the median value sent through the output wide register. The computed median replaces the centre pixel. When the median filtering is carried out in real time, the software implementation in general-purpose processors usually take long time to give results. Even though sorting operation is not so efficient on the configurable processors, implementation of median filters on them reduces the execution time.

The implementation of median filter is shown in the following figures Fig. 3 to Fig. 5.

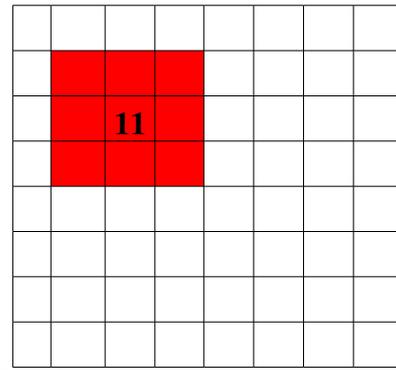

Fig. 5 Centre pixel replaced with median value

Median filter and its modification have shown good efficiency in suppressing impulse noise and capability of preserving image edges [8]. The usual median filter removes fine details in the image, which results in blurred image. Sicilian median selection algorithm is one of the divide and conquers technique [13], which is used for sorting the pixel values. But here there is no guarantee that it results in exact median value. Whereas the adaptive median filter [2] alters only the noisy pixels and it preserves the edges [7].

The experimental results of the median filter algorithm are shown in the following figures from Fig 6 to Fig.10. The adaptive median filter algorithm is found to replace the noise completely and it best suits for the image processing applications.

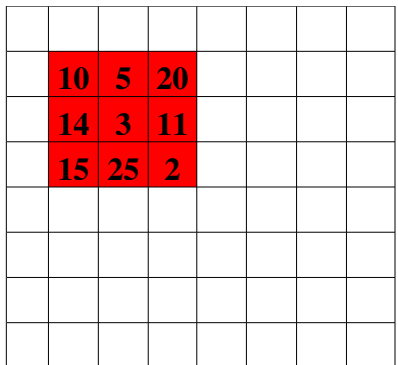

Fig. 3 3x3 sliding window

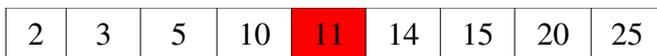

Fig. 4 Median value after sorting

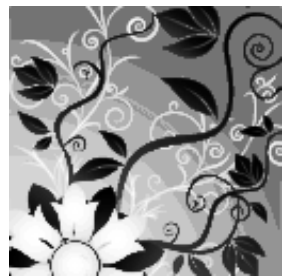 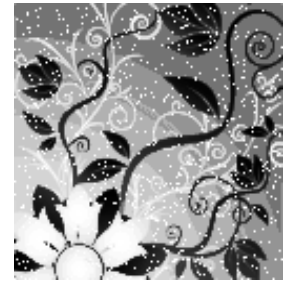

Fig. 6 3x3 Original image    Fig. 7 Noise introduced

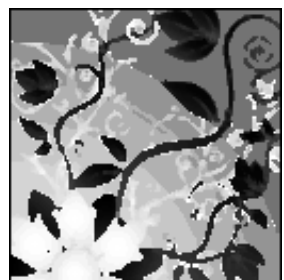 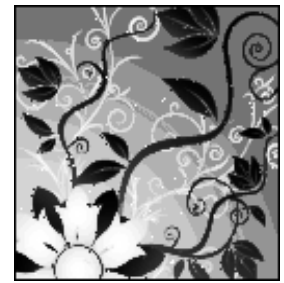

Fig. 8 Median filtering    Fig. 9 Sicilian median

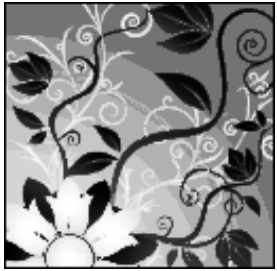

Fig. 10 Adaptive median filter

The DVR technology designed for video surveillance industry uses latest SCPs, is a specialised digital microprocessor used to efficiently and rapidly perform calculations on digitised signals that were originally analog in form, such as audio and video. The big advantage of SCP lies in the programmability of the processor, allowing parameters to be easily changed. The DVR hardware part (DVR board) consists of one or more SCPs.

The DVR board [6], which is responsible for encoding, decoding can also, do frame analysis to verify motion, night, blind and privacy regions. This technology supports latest compression technologies for encoding. The development kit is responsible for application level programming interface in order to communicate with the hardware. The median filtering algorithms are implemented on digital video recorder by grabbing the frames from the host side. Thirty frames are processed per second in real time from a single channel. These frames are dependent on the type of video standard used (NTSC/PAL).

Figures 11 and 12 show the experimental results of the median filter algorithm run on the DVR board. The single image contains two parts. The first part which contains noise and second part is the noise free image which is obtained after applying adaptive median filter algorithm on it.

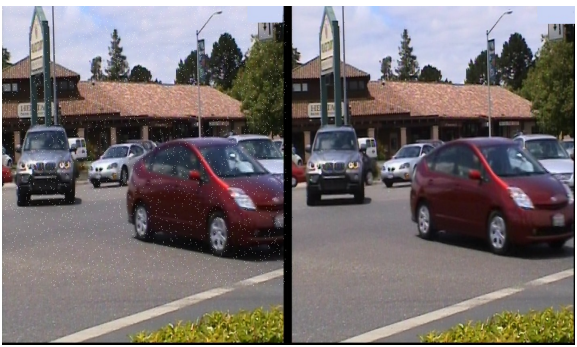

Fig. 11 Adaptive Median filtering on DVR

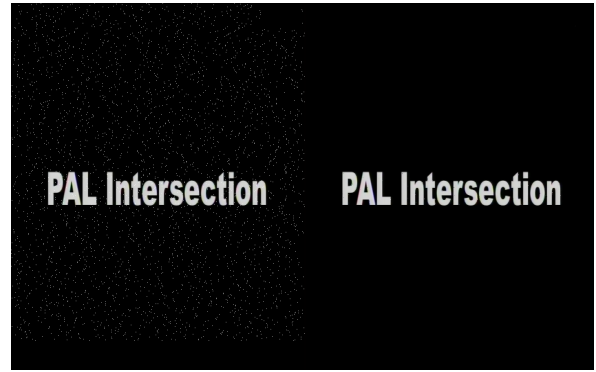

Fig. 12 Adaptive Median filtering on DVR

## IV. APPLICATION OF SCPA

Software Configurable Processor Array (SCPA) is a group of S6 processors. With the help of high speed hardware interface it is possible to connect multiple processors to form a single compute resource and it is most suitable for parallel applications [3]. The master processor is connected to the host computer with PCI express. The multiprocessor environment provides the unique architecture for running the user applications.

SCPA provides runtime environment and communication tool kit to help build applications for processor array [3]. This environment provides co-operative multitasking, task creation, distributed memory allocation and other management services. It is also possible to exchange both bulk of data and small messages. The development environment provides to build and run the application as a native target which imitates the behaviour of the processor array. Later this can be directly run on the developmental board.

The working procedure of SCPA [3] is described as follows. The list of tasks are defined in the table format and when the runtime initialisation takes place, the first task which is defined in the table is started by PE0 , the master processor. The responsibility of SCPA run time includes initialising the scheduler, cooperative multitasking, the memory manager, synchronisation, the profiler etc. In the real time application digital video recorder does the encoding and decoding functions with the help of multiple processor communication. For a multi-channel encoder/decoder the processing is distributed to different PEs.

In order to achieve high throughput colour conversion [14], different colour conversion algorithms are run on different processing elements of the SCPA. The master processor (PE0) allocates the different colour conversion tasks to other processors. The three processors PE1,PE2,PE3 run RGB to

YCC,RGB to YIQ,RGB to CMY colour conversion [14],[11] algorithms respectively.

Host controls the master processor through PCIe ((Peripheral Component Interconnect Express) interface. Each of the processor performs these conversions and returns the results to master processor. Proper synchronisation is achieved with the help of different application programming interfaces. Figure 13 depicts the multi-processor communication between the processors, each of these processor runs one colour conversion algorithm.

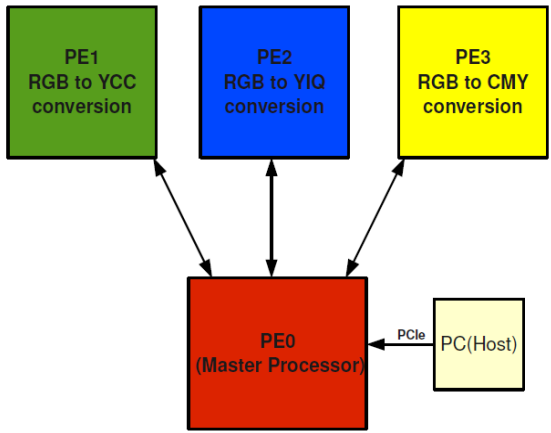

Fig. 13 Multiprocessor communication

The performance of the different colour conversion in pixel/compute-cycle is listed in the Table. 1 and 2. The first table shows the performance results of multiple applications run on the developmental board. Here each application runs on the different processor. The second table shows the performances of multiple processors communication among the processors by using application programming interfaces like send, receive etc. Here master processor allocates application to different processors and collect the computed results back. In the following Table.1 and Table.2 due to the shared memory and co-operative multitasking, the performance deteriorates the PEs other than PE0.

Table. 1 Performance of multiple PEs

| Application Name | Performance |
|---|---|
| RGB to YIQ | 1.28 |
| RGB to YCC | 1.96 |
| RGB to YUV | 2.32 |
| RGB to CMY | 2.52 |

Table. 2 Performance of inter-process communication

| Application Name | Performance |
|---|---|
| RGB to YCC | 1.65 |
| RGB to CMY | 1.84 |
| RGB to YIQ | 1.95 |

## V. CONCLUSIONS

We have presented the results of implementation of median filtering on the software configurable processor. The hardware acceleration feature of the SCP namely, to run compute-intensive parts on the ISEF as extension instructions are exploited to enhance the performance of the algorithm. We achieved an acceleration factor of 5 over non ISEF based implementation. The same algorithm is also applied on the video frames coming from digital video recorder in real time, thus proving the real time performance of SCP.

We have presented the results of running various colour conversion algorithms on the SCP array. The inter-process communication between the processors is used to pass to and fro the colour pixel data between processors. Thus, the software configurable processor provides designer friendly coding environment in low cost and best suits for accelerating embedded system applications.

### Acknowledgment
We thank Stretch Inc USA and Manipal Dot Net Pvt. Ltd., for providing the opportunity to write this paper.